\documentclass[12pt]{article}
\usepackage{graphics,amsfonts,amsmath,times,parskip,apalike,epsf,epsfig,latexsym,rotating}
\usepackage{amssymb,amsthm,amsbsy}
\def\bbbc{{\mathchoice {\setbox0=\hbox{$\displaystyle\rm C$}\hbox{\hbox
to0pt{\kern0.4\wd0\vrule height0.9\ht0\hss}\box0}}
{\setbox0=\hbox{$\textstyle\rm C$}\hbox{\hbox
to0pt{\kern0.4\wd0\vrule height0.9\ht0\hss}\box0}}
{\setbox0=\hbox{$\scriptstyle\rm C$}\hbox{\hbox
to0pt{\kern0.4\wd0\vrule height0.9\ht0\hss}\box0}}
{\setbox0=\hbox{$\scriptscriptstyle\rm C$}\hbox{\hbox
to0pt{\kern0.4\wd0\vrule height0.9\ht0\hss}\box0}}}}

\title{Bayesian matching of unlabelled point sets using Procrustes and configuration models}
\author{Kim Kenobi,\\Centre for Plant Integrative Biology, University of Nottingham, UK\\ 
and\\Ian L.~Dryden\footnote{Email correspondence: {\tt ian.dryden@nottingham.ac.uk}},\\
School of Mathematical Sciences,
University of Nottingham, UK.}

\date{}

\begin{document}
\bibliographystyle{apalike}
\maketitle

\begin{abstract}
\setlength{\parindent}{0cm}

The problem of matching unlabelled point sets using Bayesian inference is considered. 
Two recently proposed models for the likelihood are compared, based on the Procrustes
size-and-shape and the full configuration. 
Bayesian inference is carried out for matching point sets using 
Markov chain Monte Carlo simulation.  An improvement to the existing Procrustes algorithm is proposed which improves 
convergence rates, using occasional large jumps in the burn-in period. 
The Procrustes and configuration methods are compared in 
a simulation study and using real data, where it is of interest to estimate the 
strengths of matches between protein binding sites. The performance of both methods
is generally quite similar, and a connection between the two models is made
using a Laplace approximation.
\end{abstract}


\section{Introduction}
Matching configurations of points is an important but challenging problem 
in many application areas, including in bioinformatics and computer vision. 
In this paper we compare two Bayesian approaches that have
been developed for matching unlabelled point sets. The matching problem, where the sets
of points may be of different sizes, is relevant for the comparison of
molecules and the comparison of objects from different views in computer vision. 
For example, if we have two protein surfaces, a question of interest
is whether the two surfaces have a region of the same shape.  This region may
correspond to a binding site that the proteins have in common; for example they
may both bind to the same protein molecule.

In this paper we compare and build on 
on the Markov chain Monte Carlo (MCMC) methods
recently independently developed by \nocite{Drydenetal07}
Green and Mardia (2006), Dryden et al. (2007) and Schmidler (2007), 
which themselves have connections with work stemming from Moss and Hancock (1996)
and Rangarajan et al. (1997), among others. 
\nocite{Mosshanc96}
\nocite{Rangarajanetal97}  

 Green and Mardia (2006) include 
 details of a small dataset where the problem is one of matching
 unlabelled point sets, and we use this dataset as a testbed for our comparisons. 
The dataset consists of the coordinates of the centres
 of gravity of the amino acids that make up the nicotinamide adenine dinucleotide phosphate (NADP) 
binding sites of two
 proteins.  Protein 1 is the human protein 17-beta hydroxysteroid
 dehydrogenase.  Protein 2 is the mouse protein carbonyl reductase. 
 The active site of protein 1 contains 40 amino acids and the active site of
 protein 2 contains 63 amino acids.  
Green and Mardia (2006) implemented their MCMC algorithm on the 
protein data. 
In Table 4 of Green and Mardia (2004)  \nocite{Greemard04} (which is not in Green and Mardia, 2006) 
each of the suggested pairings between
amino acids in protein 1 and in protein 2 is assigned a probability.  These probabilities 
were estimated by observing how often those matches were represented in long
runs of the MCMC algorithm after convergence, and we use these findings as a basis for comparing 
the algorithms.

This paper consists of two main contributions.  First we describe an improvement to the 
algorithm of Dryden et al. (2007) to prevent it from getting trapped in local modes in the burn-in 
period.  This method involves
introducing some irreversible big jumps to find a good starting point for the
MCMC algorithm.  Secondly we compare the performance of MCMC algorithms for simulating from two different
Bayesian models: involving Procrustes matching (as in Dryden et al., 2007 and Schmidler, 2007) and involving the full 
configuration (as in Green and Mardia, 2006).

\section{Procrustes model}
\subsection{Match matrix}
Consider two configurations of $M$ and $N$ points in $m$ dimensions, 
and we write $X$ as an $M \times m$ matrix and $\mu$ as a $N \times m$ matrix 
of co-ordinates. In our application the configurations are molecules, the
points are amino acid functional site centroids, and the configurations are 
in $m = 3$ dimensions. A key part of protein molecule matching is to identify 
which functional sites correspond between two molecules. In chemoinformatics 
when comparing smaller drug molecules the points are atoms and it is of interest 
to find correspondences between pairs of atoms in molecules. 

In order to specify the labelling or correspondence between the points we use a 
\emph{match matrix} $\Lambda$, which is a $M \times (N+1)$ matrix 
of 1s and 0s, in which every row sums to 1 to
represent a particular matching of the points in $X$ to the points in $\mu$.  For
$1\leq j \leq N$, if $\lambda_{ij}=1$ then the $i^{th}$ point of $X$ matches to
the $j^{th}$ point in $\mu$.  If $\lambda_{i,N+1}=1$ then the $i^{th}$ point of
$X$ does not match to any point in $\mu$. Note that there is no requirement for the
columns to sum to 1, and so many-to-one matches are allowed. Also, the matching is 
not symmetric - in general the match from point set A to B will differ from 
the match from point set B to A. 

We shall consider two approaches to molecule matching using different Bayesian models: 
a Procrustes size-and-shape model (Dryden et al., 2007; Schmidler, 2007) and a 
configuration model (Green and Mardia, 2006). The methods use
Markov Chain Monte Carlo (MCMC) simulation to draw inferences about the match
matrix and a concentration parameter, although the treatment of the 
rotation and translation nuisance parameters differs.

\subsection{Likelihood}
Given a match matrix, $\Lambda$,
with $p$ matching points, and configuration matrices 
$X$ and $\mu$ which we assume have been centred, let
$X^{\Lambda}$ be a $p \times m$ matrix of the rows of $X$ for which $\lambda_{i,N+1}=0$ (i.e. the
matched points in $X$).  Let $\mu^{\Lambda}$ be a $p \times m$ matrix of the rows of $\mu$ which
correspond to the points in $\mu$ to which the points of $X^{\Lambda}$ are
matched. We regard $X$ as a random configuration and $\mu$ as fixed. 

A rotation of $X$ is given by post-multiplication by a rotation matrix $\Gamma \in SO(m)$, 
where $SO(m)$ is the special orthogonal group of $m \times m$ matrices such that $\Gamma^T \Gamma = \Gamma \Gamma^T = I_m$ 
and $| \Gamma | = 1$.  A translation of $X$ is given by addition of each row by $\gamma^T \in \mathbb{R}^m$. 
The size-and-shape of the configuration consists of all geometrical 
properties that are invariant under rotation and translation of $X^{\Lambda}$, i.e. 
 the size-and-shape of  $X^{\Lambda} \Gamma + 1_p \gamma^T$ is the same as that of $X^{\Lambda}$ 
(see Dryden and Mardia, 1992; 1998, Chapter 8).  Here $1_p$ is the $p$-vector of ones, and $I_m$ is the 
$m \times m$ identity matrix. 

We first use partial Procrustes registration to register $X^{\Lambda}$ to
$\mu^{\Lambda}$, in order to define a distance between the size-and-shapes. This aspect of the matching is 
present in both the Dryden et al. (2007) and Schmidler (2007) approaches. 
The Procrustes matching involves finding $\hat{\Gamma}\in SO(m)$ and
$\hat{\gamma}\in \mathbb{R}^m$ such that 
$$\parallel\mu^{\Lambda}-X^{\Lambda}\hat{\Gamma}-1_p\hat{\gamma}^T\parallel=\mathop{\inf_{\Gamma\in
SO(m)}}_{\gamma\in\mathbb{R}^p}
\parallel\mu^{\Lambda}-X^{\Lambda}\Gamma-1_p\gamma^T\parallel=d_S(X^{\Lambda},\mu^{\Lambda}),$$
where $d_S(X^{\Lambda},\mu^{\Lambda})$ is the Riemannian metric in
size-and-shape space, $S\Sigma_m^p$  (see Kendall, 1989; Dryden and Mardia, 1992, 1998). \nocite{Drydmard98,Drydmard92,Kendall89}
The Procrustes estimators of rotation and translation, $\hat{\Gamma}$ and
$\hat{\gamma}$ are
$$\hat{\gamma}=0_p, \hat{\Gamma}=R_1R_2^T,\hspace{1in}R_1,R_2\in SO(m),$$
where $(\mu^{\Lambda})^T X^{\Lambda}=
\| X^{\Lambda} \| \| \mu^{\Lambda} \| R_2DR_1^T$ and
$D=\mathrm{diag}(l_1,l_2,\ldots,l_m)$ is an $m\times m$ diagonal matrix 
where the
eigenvalues, $l_j$, are optimally signed $(l_1\geq l_2\geq \ldots \geq|l_m|\geq 0)$ and
non-degenerate $(l_{m-1}+l_m>0)$, see Kent and Mardia (2001). \nocite{Kentmard01}

Let $\hat{X}^{\Lambda}=X^{\Lambda}\hat{\Gamma}+1_p\hat{\gamma}^T.$  Then $\hat{X}^{\Lambda}$ is the partial Procrustes fit of $X^{\Lambda}$ onto
$\mu^{\Lambda}$.  (It is `partial' because no scaling has been used, just
rotation and translation.)

The partial Procrustes tangent coordinates of $X^{\Lambda}$ at $\mu^{\Lambda}$
are given by the $p \times m$ matrix 
$$V^{\Lambda}=\hat{X}^{\Lambda}-\mu^{\Lambda}=X^{\Lambda}\hat{\Gamma}+1_p\hat{\gamma}^T-\mu^{\Lambda}$$
which is in a $pm-m(m-1)/2-m$ dimensional linear subspace of $\mathbb{R}^{mp}$.

We denote the unmatched points in $X$ by $X^{-\Lambda}$.  We transform these
points using the same transformation parameters as for $X^{\Lambda}$.  Let
$\hat{X}^{-\Lambda}=X^{-\Lambda}\hat{\Gamma}+1_{M-p}\hat{\gamma}^T$.  We
consider $X^{-\Lambda}$ to lie in $\mathbb{R}^{(M-p)m}$.

Given the match matrix, $\Lambda$, the size-and-shape of $X^{\Lambda}$ lies in 
$S\Sigma_m^p$ and $X^{-\Lambda}$ lies in $\mathbb{R}^{(M-p)m}$.

We assume a zero mean isotropic Gaussian model for $V^{\Lambda}$ in $Q=pm-m(m-1)/2-m$
dimensions.  (There are $m(m-1)/2+m$ linear constraints on $V^{\Lambda}$ due to the
Procrustes registration.)  We assume that $X^{-\Lambda}$, the non-matching part,
is uniformly distributed in a bounded region, $\mathcal{A}$, with volume
$|\mathcal{A}|$ of $\mathbb{R}^m$. For the protein data we use $|\mathcal{A}| = 25500$ which is 
the volume of a bounding box obtained by multiplying the maximum lengths in the $x,y,z$ directions 
for each protein. 

The likelihood of $X$ given $\Lambda$ and $\tau=1/\sigma^2$, a precision
parameter where $\sigma^2$ is a measure of the variability at each point, is
\begin{eqnarray*}
L(X|\Lambda,\tau,\mu)&=&f_{V^{\Lambda}}(V^{\Lambda}|\tau,\Lambda,\mu)f_{X^{-\Lambda}}(X^{-\Lambda}|\Lambda)\\
&=&(2\pi)^{-Q/2}\tau^{Q/2}\exp\left( -\frac{\tau}{2}\mathrm{trace}\{(V^{\Lambda})^TV^{\Lambda}\}\right)\times\frac{1}{|\mathcal{A}|^{M-p}}\\
&=&(2\pi)^{-Q/2}\tau^{Q/2}\exp\left( -\frac{\tau}{2}d_S(X^{\Lambda},\mu^{\Lambda})^2\right)\times\frac{1}{|\mathcal{A}|^{M-p}}.\end{eqnarray*}
This likelihood is given by Dryden et al. (2007) and essentially is that of Schmidler (2007) (with $Q=mp$ in the latter). 
\subsection{Prior and posterior distributions}\label{prior1}
We write $\pi(\tau)$ and $\pi(\Lambda)$ for the prior distributions of $\tau$
and $\Lambda$ and assume $\tau$ and $\Lambda$ are independent \emph{a priori}.
We use the prior distribution $\tau\sim \Gamma(\alpha_0,\beta_0)$.

  For the prior 
distribution of 
$\Lambda$, we assume the rows are independently distributed with the $i^{th}$
row having distribution
$$\pi(\lambda_{i,N+1}=1)=\psi, \; \; \pi(\lambda_{ij}=1)=\frac{1-\psi}{N},\hspace{1cm}1\leq
j\leq N,$$
for $1\leq i \leq M$ and $0\leq\psi\leq 1$.
If $\psi=\frac{1}{N+1}$ then $\Lambda$ is uniformly distributed in
$\mathcal{M}_{M,N+1}$, the space of possible $M \times N+1$ match matrices.
The posterior density of $\tau$ and $\Lambda$ conditional on $X$ is 
$$\pi(\tau,\Lambda|X,\mu)=\frac{\pi(\tau)\pi(\Lambda)L(X|\Lambda,\tau,\mu)}{\sum_{\Lambda}\int_0^{\infty}\pi(\tau)\pi(\Lambda)L(X|\Lambda,\tau,\mu)\mathrm{d}\tau}. $$

\subsection{MCMC Inference}
The full conditional distribution of $\tau$ is available from the conjugacy of
the Gamma distribution,
$$(\tau|X,\Lambda,\mu)\sim\Gamma\left(\alpha_0+\frac{Q}{2},\beta_0+\frac{d_S(X^{\Lambda},\mu^{\Lambda})^2}{2}\right) , $$
so we update $\tau$ with a Gibbs step.

We make updates to the match matrix using a Metropolis-Hastings step.  We select
a row at random and move the $1$ to a new position in $[1,\ldots,N+1]$. 
In particular, if the selected point is already matched then it becomes unmatched with 
probability $p_{reject}$, or it is matched to another point $i$ with 
probability $(1-p_{reject})/(N-1)$. If the selected point is unmatched then it becomes 
matched to point $i$ with probability $1/N$. 

We
accept the new proposal, $\Lambda^*,$ with probability
$$\alpha_{\Lambda}=\mathrm{min}\left\{1,\frac{\pi(\Lambda^*|X,\mu,\tau) q}{\pi(\Lambda|X,\mu,\tau) q^*}\right\}, $$
where 
$$q/q^*=\left\{\begin{array}{cl}
p_{reject}/(1/N) & \mbox{if we are making an unmatched point matched}\\
(1/N)/p_{reject} & \mbox{if we are making a matched point unmatched}\\
1 & \mbox{if we are making a matched point match a different point in protein
2.} \end{array}\right.$$
If $p_{reject} = 1/N$ then $q/q^* = 1$, which was the value used by Dryden et al. (2007).

Dryden et al.(2007) also describe a computationally faster approximate Metropolis-Hastings update
to the match matrix which does not require the use of the whole configuration in
the calculation of the density.  If we propose the change $(i\rightarrow l_1)$
to $(i\rightarrow l_2)$ then the alternative Hastings ratio,
$\alpha^*_{\Lambda}$ is given by
\begin{equation}
\label{alpha.star.eqn}
\alpha^*_{\lambda}=\mathrm{min}\{g(x_i,\mu_{l_2})q/(g(x_i,\mu_{l_1})q^*),1\},
\end{equation}
where 
$$g(x_i,\mu_j)=\left\{\begin{array}{cl}
\frac{1-\psi}{N}\left(\frac{\tau}{2\pi}\right)^{m/2}\exp\left(-\frac{\tau}{2}|x_i-\mu_j|^2\right),
& \mbox{if $j<N+1$}\\
\psi\frac{1}{|\mathcal{A}|}& \mbox{if $j=N+1.$}\end{array}\right.$$
When a new match is accepted the ordinary partial Procrustes registration is
carried out on the new matching points to ensure the configuration of matching
points has rotation removed. 

For brevity we shall refer to the size-and-shape model as the ``Procrustes model", 
and matching using MCMC simulation with this model as the ``Procrustes method". 
Note that Schmidler (2007) uses geometric hashing for computationally fast approximate inference, 
which we do not consider here.  

\subsection{Improving the Procrustes algorithm}
One of the problems with the MCMC scheme is
that because of the multimodality of the likelihood function for the match
matrix $\Lambda$, the molecules often get stuck in a local mode. 
In order to circumvent this problem Dryden et al. (2007) ran the algorithm from 
a number of different start points until the algorithm had reached 
a position which satisfied certain convergence criteria. 

We propose a new initialisation algorithm which involves proposing 
much more radical changes to the match matrix than changing just one row.  The four types of bigger 
moves are called `nearness',`rotation',`translation' and `flip'.   All four types of proposal
are non reversible, and therefore we only allow these big jumps at the start of the
MCMC algorithm.  Effectively the use of these proposals helps to find a good
starting point for the subsequent MCMC inference. The new moves are: 

\begin{enumerate}
\item {\bf Nearness.} Each of the matched points in $X$ (i.e. those rows
of $\Lambda$ that have a 0 in the last column) is matched to the point in $\mu$
that is nearest to it.
Let $I^{\Lambda}$ be the index of matched points, so
$I^{\Lambda}=\{i\in \{1,2,\ldots,M\}:\lambda_{i,N+1}=0\}$. 
We define $\Lambda^{*}=(\lambda_{ij}^{*})$ by
$$\lambda_{ij}^{*}=\left\{\begin{array}{cl}1&i\notin I^{\Lambda},j=N+1\\
1&i\in I^{\lambda},\parallel (X)_i-(\mu)_j\parallel=\min_{l\in\{1,\ldots,N\}}
\parallel (X)_i-(\mu)_l\parallel \\ 0 &
\mbox{otherwise.}\end{array}\right.$$
Let $N(X,\mu,\Lambda)=\Lambda^{*}$ as defined above.  Note that 
$\Lambda^{*}$ has the same number of matched points as $\Lambda$.
The other three methods (rotation, translation and flip) use this nearness
step at the end.

\item {\bf Rotation.} 
Randomly choose an angle $\theta\sim U[-\pi,\pi]$.  Randomly choose an axis
($x,y$ or $z$) about which to rotate, and   
set $R=R_x(\theta),R_y(\theta)$ or $R_z(\theta)$ as appropriate, where $R_x,R_y,R_z$ are 
defined in (\ref{RxRy}) and (\ref{Rz}). 
Let $X^{*}=XR$ then map each point in $X^{*}$ to the nearest
point in $\mu$, i.e.
$\Lambda^{*}\equiv R^{\Lambda}(X,\mu,\Lambda)=N(X^{*},\mu,\Lambda)$.

\item {\bf Translation.}
Choose $\gamma\sim N_3(0,\sigma^2)$.  Define $X^{*}=X+1_{M}\gamma^T$ and then map
each point in $X^{*}$ to its nearest point in $\mu$.  Thus 
$\Lambda^{*}\equiv T^{\Lambda}(X,\mu,\Lambda)=N(X^{*},\mu,\Lambda)$. 

\item{\bf Flip.}
This move has the same form as the rotation step, but instead of selecting $\theta$
from a $U[-\pi,\pi]$ distribution we set $\theta=\pi$.  
\end{enumerate}

We define an initialisation phase by setting a maximum number of initial jumps,
$N_{initialisation}$.  We also define a settling time, $N_{settle}$.  During 
the initialisation phase (i.e. $ \le N_{initialisation}$ interactions)
 at least 
$N_{settle}$ default updates are proposed between any two big jump
proposals.  The rationale behind this is to explore the region of the parameter
space we `land in' after making a big jump before immediately jumping somewhere
else.  The hope is that the settling time allows the algorithm to home in on a
solution if a big jump takes us somewhere close to the optimal solution. 
Provided at least $N_{settle}$ default updates have been proposed we randomly
choose an update type from $\{\mbox{nearness, rotation, translation, flip, default}\}$,
with probabilities $p_n,p_r,p_t,p_f,1-(p_n+p_r+p_t+p_f)$, say.  Whichever update
method is chosen, a new match matrix, $\Lambda^{*}$ is generated.  We then
accept the new match matrix with probability
$$\alpha_{\Lambda}=\min\{1,\pi(\Lambda^{*}|X,\mu,\tau)/\pi(\Lambda|X,\mu,\tau)\}.$$

After $N_{initialisation}$ iterations the algorithm proceeds exactly as
described in Dryden et al. (2007).

%

\section{Configuration model}
\label{Configuration.sec}
\subsection{Likelihood}
We now consider an alternative model for the configuration of points which 
turns out to be equivalent to that of Green and Mardia (2006). 
We again assume that $\mu$ is a fixed $N\times m$ configuration and $X$ is an $M \times
m$ configuration that we apply rigid-body transformations to.  

This model for the co-ordinates of the points does not involve 
removing rotation and translation by Procrustes matching. Rather, the rotation matrix 
$\Gamma \in SO(m)$ and the translation parameter $\gamma$ will be parameters in the model. 
The matched points in $X^{\Lambda}$ are taken as Gaussian perturbations of the matching points in $\mu$, 
and we assume that the rows of 
$X^{-\Lambda}$ are distributed uniformly over
a bounded region $\mathcal{A}\subset\mathbb{R}^m$ of volume $|\mathcal{A}|$. We concentrate on the $m=3$ 
dimensional case here.

Given an $M\times(N+1)$ match matrix, $\Lambda$ (with $p$ matching points), 
rotation matrix $\Gamma$ and translation vector $\gamma$ the likelihood is
therefore defined as: 
$$L^*(X|\Lambda,\mu,\tau,\Gamma,\gamma)=\left(\frac{1}{2\pi}\right)^{3p/2}\tau^{3p/2}\exp\left(-\frac{\tau}{2}\mathrm{trace}\{(\tilde{X}^{\Lambda}-\mu^{\Lambda})^T(\tilde{X}^{\Lambda}-\mu^{\Lambda})\}\right)\times\frac{1}{|\mathcal{A}|^{M-p}},$$
where $\tilde{X}^{\Lambda}=X^{\Lambda} \Gamma + 1_p \gamma^T$,  
$\Gamma = R_z(\theta_{12}) R_y(\theta_{13}) R_x(\theta_{23})$, and 
the rotation matrices about the $x,y,z$ axes are: 
\begin{equation}
R_x(\theta_{23}) = 
 \left(\begin{array}{ccc}
1 & 0 & 0\\
0 & \cos\theta_{23}&\sin\theta_{23}\\
0 & -\sin\theta_{23}&\cos\theta_{23}
\end{array}\right)  \; \; , \; \; 
R_y(\theta_{13}) =    \left(\begin{array}{ccc}
\cos\theta_{13}&0&\sin\theta_{13}\\
0 & 1 & 0 \\
-\sin\theta_{13}&0&\cos\theta_{13}
\end{array}\right) , \label{RxRy} 
\end{equation}
\begin{equation}
R_z(\theta_{12}) =   \left(\begin{array}{ccc}
\cos\theta_{12}&\sin\theta_{12}&0\\
-\sin\theta_{12}&\cos\theta_{12}&0\\
0&0&1\end{array}\right) ,  \label{Rz}
\end{equation}
with Euler angles $\theta_{12} \in [-\pi,\pi), \theta_{13} \in [-\pi/2,\pi/2], \theta_{23} \in [-\pi,\pi)$. There are many choices of 
Euler angle representations and all have singularities (Stuelpnagel, 1964), although the singularities 
have measure zero 
with respect to Haar measure which is given by  
\nocite{Stuelpnagel64}
$$ \frac{1}{8 \pi^2} \cos(\theta_{13}) d\theta_{12}d\theta_{13}d\theta_{23}  $$
in this case (e.g. see Khatri and Mardia, 1977). \nocite{Khatmard77}

Note that Green and Mardia (2006)'s model is constructed with $X$ and $\mu$ as Gaussian perturbations 
from an underlying Poisson process. However, the likelihood is actually of the same form as the one sided
version, where $X$ is perturbed from $\mu$, although the variance parameter is doubled. 

\subsection{Prior and posterior distributions}
We take $\tau, \Lambda, \Gamma, \gamma$ to be mutually independent {\it a priori}, and the 
priors of $\tau$ and $\Lambda$ are taken as in Section \ref{prior1}. 
We also take the prior for $\gamma$ as: 
$$\gamma \sim N_3(\mu_{\gamma},\sigma_{\gamma}^2I) , $$
and we take $\Gamma$ to be uniform with respect to Haar measure on $SO(m)$. 
The posterior density of ($\Lambda, \tau, \Gamma, \gamma)$ conditioned on
$X$ is  
$$\pi(\tau,\Lambda,\Gamma,\gamma |X,\mu)=\frac{\pi(\tau)\pi(\Lambda)\pi(\Gamma)\pi(\gamma)L(X|\Lambda,\tau,\mu,\Gamma,\gamma)}{\sum_{\Lambda}\int_0^{\infty}\pi(\tau)\pi(\Lambda)\pi(\Gamma)\pi(\gamma) L(X|\Lambda,\tau,\mu,\Gamma,\gamma)\mathrm{d}\tau}. $$

\subsection{MCMC simulation}
The full conditional distribution of $\tau$ is given by 
$$ (\tau | X , \Lambda, \Gamma, \gamma, \mu) \sim \Gamma \left(  \alpha_0 + \frac{3p}{2} , \beta_0 + 
\frac{ \|  \tilde{X}^{\Lambda} - \mu^{\Lambda} \|^2 }{2} \right) , $$
and so a Gibbs update can be used for $\tau$.

We update the rotation angles using a Metropolis-Hastings
step, drawing the proposal perturbations from a uniform distribution on 
$[-0.2,0.2]$ for $\theta_{12}, \theta_{23}$, and uniform on $[-0.1,0.1]$ for $\theta_{13}$, 
to give proposed angles $\theta_{12}^*,\theta_{13}^*,\theta_{23}^*$. 
The Hastings ratio is: 
$$\min \left( 1, \frac{ \pi(\tau,\Lambda,\Gamma(\theta_{12}^*,\theta_{13}^*,\theta_{23}^*),\gamma |X,\mu) \cos \theta_{13}^* }
{ \pi(\tau,\Lambda,\Gamma(\theta_{12},\theta_{13},\theta_{23}),\gamma |X,\mu) \cos \theta_{13} }  \right) ,$$
and the extra cosine terms are due to the Haar measure on the special orthogonal rotation group.

The full conditional distribution of $\gamma$ is given by
\begin{equation}
\label{conjugacy2.eqn}\gamma|X,\mu,\tau,\Lambda,\Gamma\sim
N\left(\frac{\mu_{\gamma}/\sigma_{\gamma}^2+\tau\sum_{j\leq M,k\leq
N,\lambda_{jk}=1}(\mu_k-x_j \Gamma)}{p\tau+1/\sigma_{\gamma}^2},\frac{1}{p\tau+1/\sigma_{\gamma}^2}I\right), \end{equation}
and so we use a Gibbs update for $\gamma$.

We update the match matrix $\Lambda$ in the same way as in the Procrustes model using
the acceptance probability
\begin{eqnarray*}
\alpha_{\Lambda}&=&\mathrm{min}\left(1,\frac{\pi(\Lambda^*|X,\mu,\tau,\Gamma,\gamma)q}{\pi(\Lambda|X,\mu,\tau,\Gamma,\gamma)q^*}\right)\\
&=&\mathrm{min}\left(1,\frac{L(X|\Lambda^*,\mu,\tau,\Gamma,\gamma)\pi(\Lambda^*)q}{L(X|\Lambda,\mu,\tau,\Gamma,\gamma)\pi(\Lambda)q^*}\right).\end{eqnarray*}
Suppose $\Lambda^*$ contains the match $(i\rightarrow l_1)$ and $\Lambda$
contains the match $(i\rightarrow l_2)$, where $l_1\neq l_2$ and the match
matrices $\Lambda^*$ and $\Lambda$ are otherwise identical.  The acceptance 
probability $\alpha_{\Lambda}$ is exactly the same as that
given in Equation (\ref{alpha.star.eqn}), i.e. the fast method of 
Dryden et al. (2007). Hence the MCMC updates of $\Lambda$  for 
the Procrustes and Configuration models are more similar than they first appear.

Note that our implementation of the MCMC simulation differs slightly from Green and 
Mardia (2006) who use a matrix Fisher conjugate prior for the rotation, and update two of the rotation angles 
with a Gibbs step. In addition, Green and Mardia (2006) ensure that the matching is 1-1
between the points, whereas we do allow the possibility of many-to-one matches.

For brevity we shall refer to this model as the ``Configuration model", 
and matching using this model as the ``Configuration method". The Configuration model has been demonstrated to 
work well in a variety of situations (see Mardia et al., 2007). \nocite{Mardiaetal07}


\subsection{Laplace approximation}
Let us consider the posterior density $\pi( \Lambda , \tau ,\Gamma,\gamma | X)$. 
Note that the rotation and translation $\Gamma, \gamma$ are nuisance parameters, and one has a choice about 
how to deal with them. In the Configuration approach one samples from the full 
joint distribution of $(\Lambda,\tau,\Gamma,\gamma |X)$ and so joint inference 
of all the parameters can be carried out. However, if $\Gamma, \gamma$ are 
considered nuisance parameters then we can integrate them out to give  
the marginal density of $(\Lambda ,\tau)$  
\begin{equation}
\pi_C(\Lambda ,\tau | X ) = \int_{\Gamma, \gamma } \pi(\Lambda, \tau , \Gamma ,\gamma | X ) d\Gamma d\gamma. \label{post1}
\end{equation}

In the Procrustes  approach the match is obtained by optimizing 
over the nuisance parameters, and so we consider the different posterior density
based on 
\begin{equation}
\pi_P( \Lambda , \tau | X ) \propto \sup_{\Gamma, \gamma}  \pi(\Lambda, \tau, \Gamma, \gamma | X ) . \label{post2}
\end{equation}
We can consider (\ref{post2}) to be an approximation to the marginal density (\ref{post1}) where the integral is
approximated using Laplace's method (Tierney and Kadane, 1986). \nocite{Tierkada86} 

From a Bayesian analysis perspective it is natural to work with 
the marginal posterior distribution (\ref{post1}). 
From a shape theory perspective the analysis should be invariant under rotations or translations of the data, 
and so a uniform prior for $\Gamma, \gamma $ in (\ref{post1}) or a distribution of the form (\ref{post2}) are 
both natural. In this paper we will explore the relative performances of the two approaches in some practical 
scenarios.

\section{Applications and simulations}
\subsection{Assessment of initialisation procedure}
Here, we use the NADP-binding site protein data 
\nocite{Greemard06} to assess the efficiency of the Procrustes algorithm, both with
and without the large jump proposals.  There are 40 centres of
gravity of amino acids for protein 1 and 63 for protein 2. Following Green and 
Mardia (2006) we take the prior hyperparameters to be $\alpha_0 = 1, \beta_0 = 36, 
\mu_\gamma = 0, \sigma_\gamma = 50$, 
and we take $\psi = 0.2$. The proposal parameters $p_{n}=0.001,p_{r}=0.02,p_{f}=0.01,p_{t}=0.09,N_{settle}=850$ 
for this application. 

\begin{figure}[htbp]
\begin{center}
\includegraphics[width=12cm,angle=270]{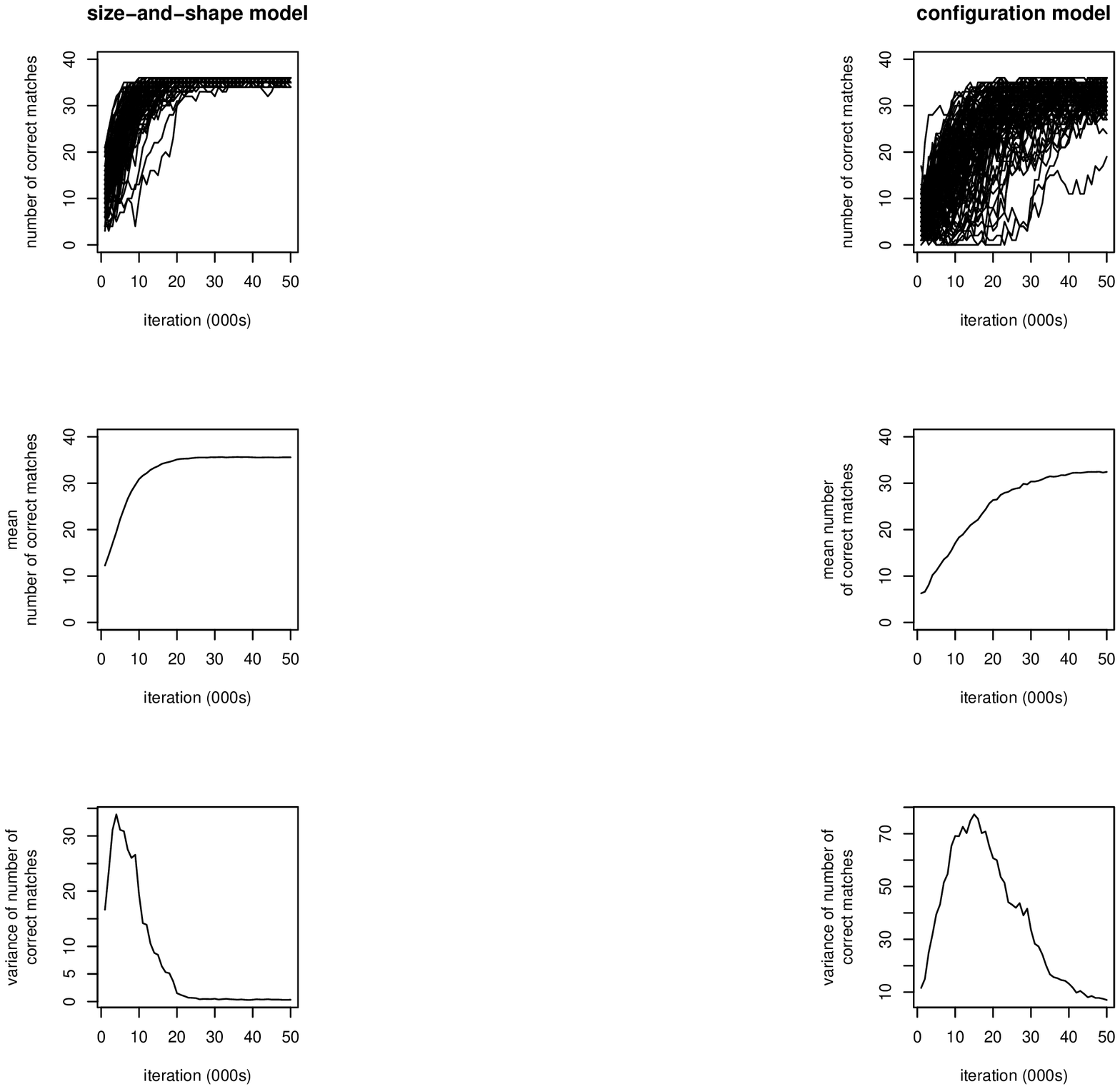}
\caption{A comparison of the numbers of correct matches over 50000 iterations (as defined in Green and
Mardia (2004) for the Procrustes and Configuration models, initialising by
choosing 10 correct matches at random}
\label{convergence.props.fig}
\end{center}
\end{figure}

We used the \emph{a priori} `correct' matches, as identified in Green and Mardia
(2004) to define a convergence criterion.  To assess the efficacy of this
criterion for determining convergence, we started 
50 MCMC runs from distinct initial
configurations in each of which 10 correct matches were selected at random.  Each run was
allowed to run for 50000 iterations, and we measured the number of correct
matches after each 1000 iterations.  The results are shown in Figure
\ref{convergence.props.fig}.  In all 50 cases, for both the Procrustes and
the Configuration models, the algorithms converged to around 36 correct matches.
 It is interesting to note that the Procrustes model converges quicker and
 more reliably, although with the large initialization proposals this is not
surprising (see the variance plots in Figure \ref{convergence.props.fig}).  
  
\begin{center}
{\bf INSERT FIGURE 1 ABOUT HERE}
\end{center}

\begin{figure}[htbp]
\begin{center}
\includegraphics[width=12cm,angle=270]{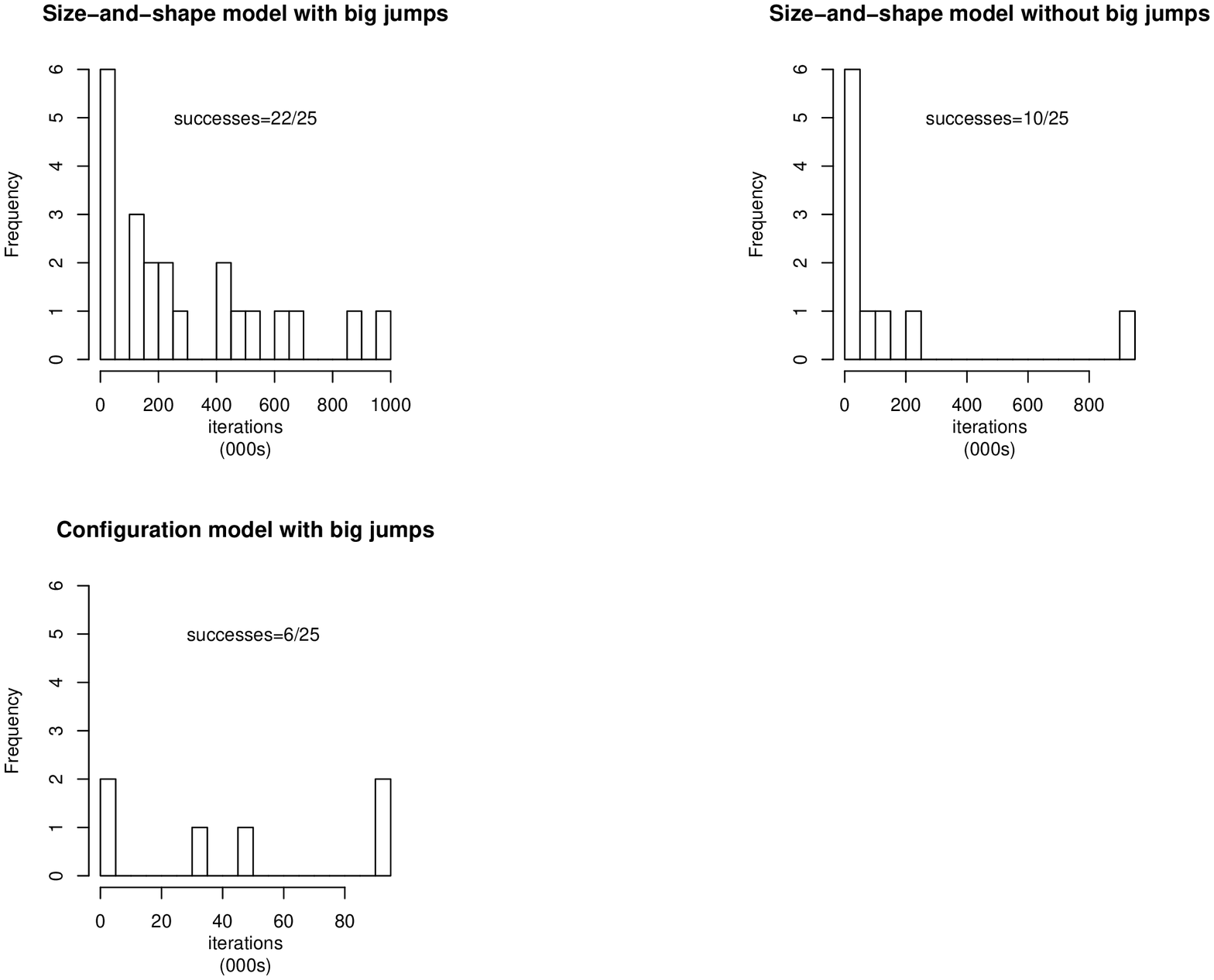}
\caption{Histograms of the number of iterations to convergence in successful
runs from random starting points for the Procrustes model, with and without
large jumps, and the Configuration model.  The parameters for the large jumps
are:
$\sigma_T=2.2,p_{n}=0.001,p_{r}=0.02,p_{f}=0.01,p_{t}=0.09,N_{settle}=850$.}
\label{convergence.hists.fig}
\end{center}
\end{figure}

To compare the convergence performance of the Procrustes and Configuration methods, we 
initiated 25 runs from random starting points.  We allowed
each run to continue for a maximum of a million iterations, monitoring the number
of correct matches after every thousand iterations.  On the basis of the results
described above, we stipulated that if within these million iterations the
number of correctly matches reached 10 then that counted as convergence.  Such
runs were allowed to continue for a further 50000 iterations.  The Procrustes method
was used both with and without the big jumps described above; these were only
used
during the initial $N_{initialisation} = 1000000$ iterations.  Figure \ref{convergence.hists.fig} shows
histograms of the number of iterations before the algorithms converged to 10
correct matches for the successful runs.  The success rates of 10/25 for the
Procrustes method without big jumps and 6/25 for the Configuration method were not too
encouraging.  However, when big jumps were included for the Procrustes method, the
success rate increased to 22/25, a very impressive result.

\begin{center}
{\bf INSERT FIGURE 2 ABOUT HERE}
\end{center}

In Green and Mardia (2006), they report convergence within a million iterations
on 83 out of 100 tests run from random starting points.  They define
convergence differently to us, looking for runs in which the log-posterior goes
higher than some threshold.  It is important to note three things when looking at
this result and comparing it with the results of Figure
\ref{convergence.hists.fig}.  Firstly, in the Green and Mardia paper, they update
the match matrix 10 times per sweep, so they are effectively looking at the
convergence within 10 million iterations.  Secondly, their proposal methods for the
angles in particular are different; they use Gibbs steps instead of
Metropolis-Hastings updates, making use of conjugacy of the matrix Fisher
distribution.  This may also improve their convergence performance, with the
form of the proposals being closer to the true distribution.  Finally, the way the model
is formulated is different, with 1-1 matches and a hidden Poisson process being used.

Although the algorithm was much more likely to converge within a million
iterations if the big jumps were included, it did mean that from certain
starting points the algorithm took a lot longer to converge if the big jumps
were included than if they were not.  This is a consequence of the choice of the
settling time parameter between large jump proposals.  One way to avoid this
might be to let the algorithms run for an initial period of 100000, say, before
introducing any big jumps.  This way, if the algorithm converged within that
period then it would not be necessary to use the big jumps at all.  
Also, the settling time between large jump proposals could be increased.
Despite the fact that it often took longer for the algorithm to converge with
the large jumps, the evidence is compelling that the big jumps vastly improve
convergence.

We experimented with the probabilities of acceptance for the four types of large
jumps.  At the levels we settled on (given in the caption of Figure
\ref{convergence.hists.fig}) the nearness proposal was always accepted (which is
always the case since the likelihood always increases for the nearness proposal),
 and the other three types were accepted roughly a quarter (flip), a third
 (rotation) and half (translation) of the
 times when they were proposed.

\subsection{Long run comparisons}
In order to compare further the Procrustes and Configuration algorithms we apply the MCMC scheme from 
a number of long runs of the method.  In order to ensure that we started the algorithms
close to convergence, we initialised the proteins by aligning the first 10
pairs of amino acids as given in Table 4 of Green and Mardia (2004). 

We ran the two algorithms and
looked at the proportion of the accepted match matrices after convergence in 
which particular matches were represented.  Although in principle many to one
matches were possible, they did not tend to occur in the long runs after
convergence.  We ran the experiment for five
values of $\psi$, the prior probability of a particular point being unmatched,
and five values of the proposal probability $p_{reject}$, the probability of moving a matched point to an
unmatched status in the proposal for the change to the match matrix.  For each
parameter the five values we used were $0.001,1/63,0.1,0.2$ and $0.4$.  (The
$1/63$ is there because $N=63$ and in the case of $\psi$, this corresponds to a
uniform prior for $\Lambda$.)

\begin{figure}[htbp]
\begin{center}
\includegraphics[width=12cm,angle=270]{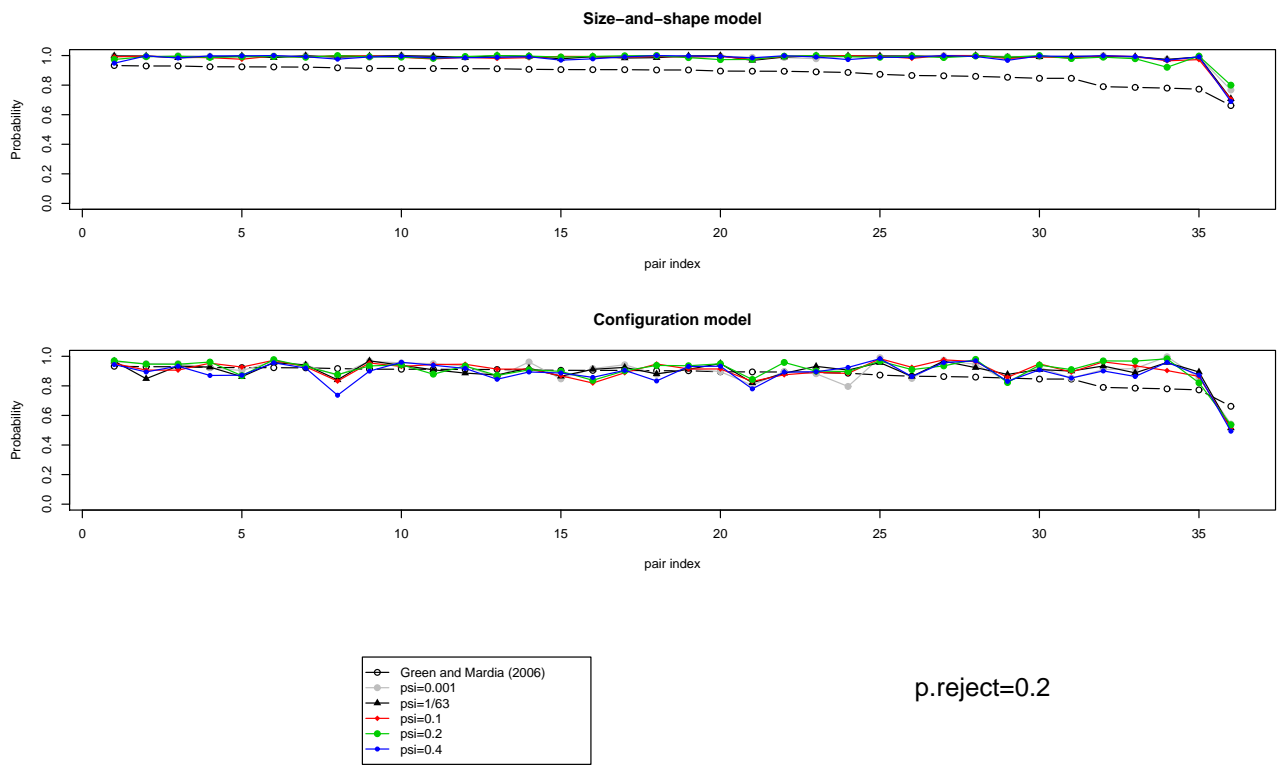}
\caption{The proportions of match matrices containing particular
pairings, based on 1000000 iterations after convergence for the Procrustes
and Configuration models for five values of $\psi$ - a comparison with the
percentages quoted in Green and Mardia (2006).}
\label{long.run.fig}
\end{center}
\end{figure}

Altering $p_{reject}$ had little effect on the results. 
We fix $p_{reject}=0.2$ and consider the effects of varying $\psi$, the prior 
probability of each point in protein 1 being unmatched (independently of the
other points).
We ran each MCMC algorithm for 1000000 iterations after convergence, adding the
match matrices together.
In Figure \ref{long.run.fig}, we show
how often the 36 most likely matches from 
Table 4 of Green and Mardia (2004) appear in our match matrices after convergence.  These
percentages are calculated as the number of times each match occurred divided by
the total number of match matrices.

\begin{center}
{\bf INSERT FIGURE 3 ABOUT HERE}
\end{center}

We have calculated
a `threshold match matrix' by putting a 1 in each position that corresponds to
the maximum entry in a row of the summed match matrices and a 0 everywhere else.
 This gives us a method for comparing how many points are matched for each value
 of $\psi$. For values of $\psi \in \{0.001,1/63,0.1,0.2,0.4\}$ the number of 
unmatched points are $\{0,0,1,4,4\}$ respectively, for both the Procrustes 
and Configuration methods. 
Clearly changing the prior distribution of $\Lambda$ by altering $\psi$ has an
effect on the number of points that are matched. 

Figure \ref{long.run.fig} shows that using the Configuration model, we obtain
probabilities for the top 36 matches reported in Green and Mardia (2006) that are
similar to the figures quoted in that paper.  However, using the Procrustes
model, the probabilities are all significantly closer to 1 than using the
Configuration model.  This suggests that the Procrustes model is `stickier'
than the Configuration model, in the sense that matches are released less 
readily after convergence.  The simulation study below investigates the
relationship of long run convergence probabilities with different variances, and
the results suggest that there is a possibility that the results observed in
Figure 
\ref{long.run.fig} may be a contingent property of the variability of the
points.  We return to this in the discussion of the simulation study.

%

Note that the posterior standard deviation $\sigma = 1/\sqrt{\tau}$ was smaller 
for the Procrustes model. In particular, the means of the 10000
values well after burn-in were 0.869 for the Procrustes model and 1.355 for the Configuration
model.  

\subsection{A simulation study}
\label{simulation.sec}

We consider now a simulation study where we know what the true probabilities of 
matching are and
compare the MCMC algorithms both with and without Procrustes registration to
see how they perform. The details of this simulation are as follows:

\newlength{\dd}
\settowidth{\dd}{Step 1}
\begin{list}{}%
{\setlength{\labelwidth}{\dd}%
\setlength{\leftmargin}{\dd}%
\addtolength{\leftmargin}{\labelsep}}
\item[\emph{Step 1}]

Define a length, $L>0$ and a minimum distance $0<d_{min}<L$.  Fix
$M,N\in\mathbb{N}$, $n_{ones}<M$.  As before, $M$ is the number of points in the
point set $X$ and $N$ is the number of points in the point set $\mu$.  
Define a vector of probabilities,
$p=(p_1,p_2,\ldots,p_M)$, where $p_1=p_2=\ldots=p_{n_{ones}}=1$ and
$p_i=0$ for $i=n_{ones}+1,\ldots,M$.  Fix $s<d_{min}$; this is the standard
deviation of the pertubations of the random points.

\item[\emph{Step 2}]

Sample the $N$ points of $\mu$ from a uniform distribution on the cube with
corners 
$$\{(-L,-L,-L),(-L,-L,L),\ldots,(L,L,L)\}$$ 
subject to the constraint that each new 
point is at least a distance $d_{min}$ from every other point.  For the
$i^{th}$ point in $X$, denoted $(X)_i$, if $p_i=1$ then we sample from a Normal
distribution centred on the $i^{th}$ point in $\mu$,
$$(X)_i\sim N_3( (\mu)_i,s^2I_3),$$
else we sample uniformly from the cube with corners as above,
$$ (X)_i \sim U[\mbox{cube as above}].$$

\item[\emph{Step 3}]

Run the two MCMC algorithms for $N_{iter}$ iterations starting from the match matrix which matches $(X)_j$
to $(\mu)_j$ for $j=1,2,\ldots,n_{ones}$.  (In other words we start the algorithms
from convergence.)  For $i=1,\ldots,,n_{ones}$, record the proportion of 
the $N_{iter}$ match matrices that match $(X)_i$ to $(\mu)_i$.  For
$i=n_{ones}+1,\ldots,M$, record the proportion of the $N_{iter}$ match matrices
for which $(X)_i$ is unmatched.

\item[\emph{Step 4}]

Hold $\mu$ constant and sample a new $X$ as described in step 2.  Repeat step 3.
 Continue this process until the proportions of successful matches and
 successfully unmatched points have been recorded for $K$ runs of the MCMC
 algorithm. 

\item[\emph{Step 5}]

Repeat experiment for various values of $s<d_{min}$.

\end{list}

\begin{figure}[htbp]
\label{MCMC.figs}
\begin{center}
\includegraphics[width=12cm,angle=270]{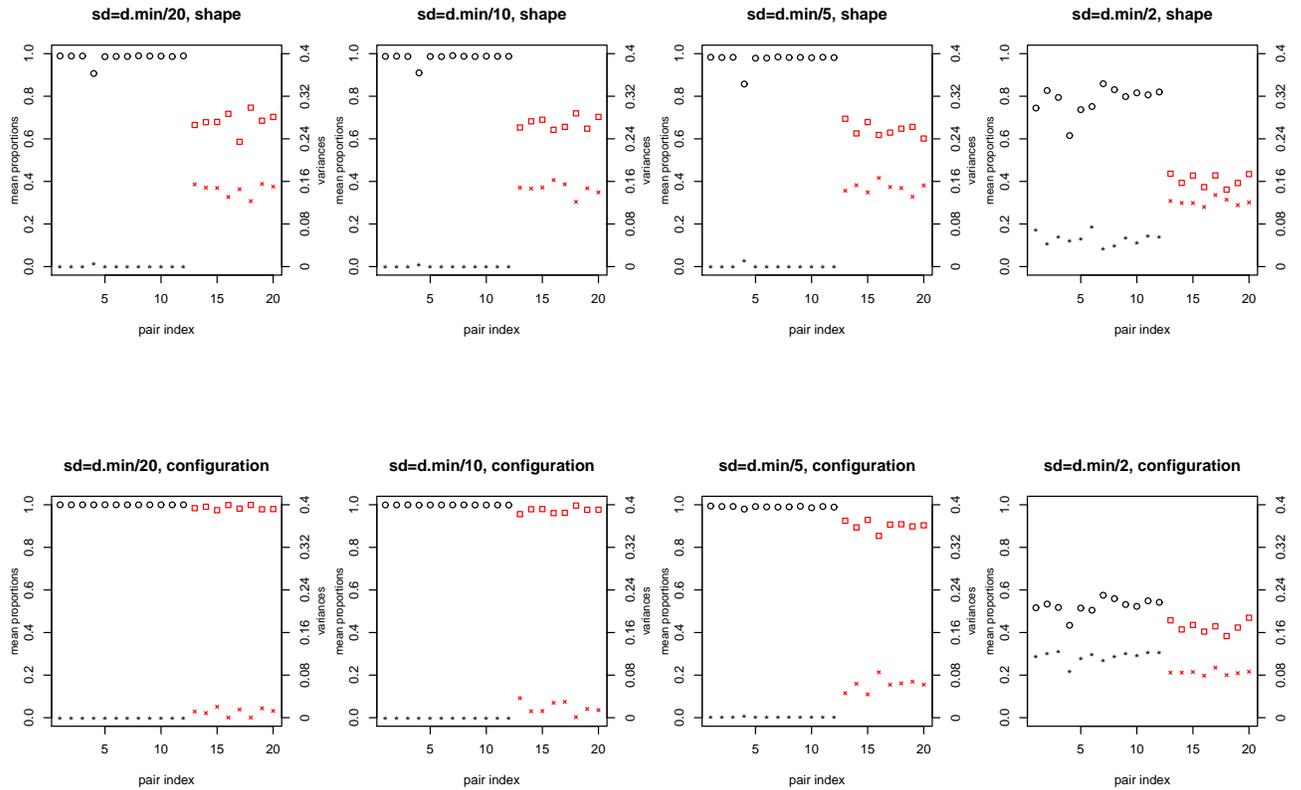}
\caption{The means (circles and squares, left hand scale) and variances (stars,
right hand scale) of the proportions of successful
matches (black circles and black small stars) and successfully unmatched points (red squares and
red large stars) with and without Procrustes
registration on long runs (100000 iterations) after convergence.  Here, there are 20
points in configuration $X$ and 24 points in configuration $\mu$.  The points in
$\mu$ are sampled uniformly from a cube of side length 20 subject to the
constraint that they are a minimum distance $d_{min}=2$ from the nearest
neighbour.  The first 12
points in $X$ are sampled from Normal distributions centred at the corresponding
points in $\mu$ and the last 8 points in $X$ are sampled uniformly on the cube
of radius $2L$.  The means and variances are calculated over 100 runs, with
$\mu$ held constant and $X$ resampled each time.}
\label{MCMC.comparison.fig}
\end{center}
\end{figure}

Figure \ref{MCMC.comparison.fig} shows the results of running this experiment
with $M=20$, $N=24$ and $n_{ones}=12$.  The values chosen for $L$ and $d_{min}$
were 10 and 2 respectively.  The experiment was run for four values of $s$, the
standard deviation parameter.  These were
$d_{min}/20,d_{min}/10,d_{min}/5,d_{min}/2$, or 0.1, 0.2, 0.4 and 1.  The value
of $N_{iter}$, the number of iterations after convergence, was 100000.

\begin{center}
{\bf INSERT FIGURE 4 ABOUT HERE}
\end{center}

Figure \ref{MCMC.comparison.fig} has a curious feature.  When the value of
the standard deviation is less than or equal to $d_{min}/5$, the Configuration
model seems to estimate the probabilities for both matched and unmatched points
more reliably than the Procrustes model.  For both models
the matched points are rarely released when the matching is very precise, but
the Configuration model gives probabilities closer to 1 than the Procrustes
model.  (This is not clear from just looking at the graphs).  When the 
standard deviation is increased to $d_{min}/2$, the Configuration
model still performs better than the Procrustes model on the unmatched
points.  Interestingly, now the Procrustes model gives significantly better
(i.e. higher) estimates for the probabilities for the matched points.

With reference to the results illustrated in Figure \ref{long.run.fig}, this
simulation study poses an interesting question.  In Figure \ref{long.run.fig},
we found that the Procrustes method appeared `stickier' than the Configuration method.  In the
light of the findings of this simulation study, it is possible that this result
is a feature of the particular relationship between the variance parameter and
the minimum distance between points in this particular dataset.  From the
simulation, it appears there may be a critical value of the standard deviation
parameter, somewhere between $d_{min}/5$ and $d_{min}/2$, for which the two MCMC
methods swap over in terms of which one gives the higher probabilities for
particular matches.  



\section{Discussion}
In conclusion, it is clear that the Procrustes method is significantly improved by 
considering the initial large jumps. However, despite quite extensive comparisons 
there is not an overall preference between the Procrustes or Configuration 
methods for all situations.  The Procrustes
method appears to converge more reliably to the true solution when the proteins are
initialised by selecting 10 correct matches at random.  This is a consequence of
the optimisation over the rotation and translation parameters that takes place
in the Procrustes method. However, for simulated datasets where the variance is small, the Configuration
method more reliably predicts the probabilities of matches, and the Procrustes method was more likely
to suffer from false matches. For larger variances the Procrustes method was more effective at 
estimating correct matches, without more false matches. In essence both models 
are fairly similar, and inference using marginal posteriors (\ref{post1}) or (\ref{post2}) is similar 
in practice due to the Laplace approximation.  

Although we have just considered pairwise matching of two configurations here, the 
methods extend to matching multiple molecules. Extensions of the Procrustes and Configuration models
for multiple alignments have been given by Dryden et al. (2007) and 
Ruffieux and Green (2008) respectively. \nocite{Ruffgree08}

The way we have set up the MCMC procedures, we do not exclude the
possibility of many-to-one matches.  We have followed the methodology of Dryden
et al. (2007) and found that in general many-to-one matches are not selected in
long runs after convergence.  However, it would be easy to constrain the choice
of match matrices such that only one-to-one matches were proposed.  This is the
method adopted by Green and Mardia (2006).


MCMC tools are an effective way of finding the
optimal correspondence and registration between two point sets where we wish to
match a subset of points from one set to a subset of points from the other set. 
But because of the combinatoric nature of looking for possible correspondences,
the algorithms are currently prohibitively time consuming for large data sets.  Suppose we were
interested in comparing two large protein surfaces to look for regions of a
similar shape (such as binding sites that are common to both proteins).  
It may be possible to use an efficient
search algorithm to scan the surface of the two proteins for small regions
that are potential candidates for binding sites and then apply the MCMC methods
to those small sites individually to confirm whether or not there are subsets of
the two regions that match well.   Schmidler (2007) \nocite{Schmidler07} notes the
difficulties of using MCMC methods for large problems and suggests the use of
geometric hashing to compute approximate posterior quantities efficiently.

\bibliography{/maths/staff/ild/tex/bibtex/fullref.bib}

\newpage

\end{document}